# Coexistence Options and Performance Analysis of 100 Gbit/s Coherent PON in Brownfield DWDM Networks

Gabriele Di Rosa[(1)], Martin Kuipers[(2)], Jim Zou[(1)], Ognjen Jovanovic[(1)], Jörg-Peter Elbers[(1)]

[(1)] Adtran Networks SE, Fraunhoferstr. 9a, 82152 Planegg, Germany, gabriele.dirosa@adtran.com
[(2)] Adtran GmbH, Hermann-Dorner-Allee 91, 12489 Berlin, Germany

**Abstract** *We study system architectures for the coexistence of future coherent PON and DWDM networks. Considering deployed optical filters, we observe filtering penalties < 1dB at a laser frequency accuracy < 12GHz when using a cost-effective architecture. ©2024 The Author(s)*

**Introduction**
The next generation of passive optical networks (PONs) beyond 50 Gbit/s may use coherent transmission techniques and provide data rates in the order of 100-200 Gbit/s. These systems are currently studied by CableLabs and ITU-T Q2/15 [1]. PONs use an Optical Line Termination (OLT) as head-end. If the Optical Network Units (ONUs) at the customer premises are located too far away from the central office, OLTs may need to be moved to outdoor locations. This can create technical (e.g. due to climatic conditions or lack of electricity) and economic (e.g. due to a low subscriber density) challenges. Substantial benefits can then be expected by distributing PON signals over existing dense wavelength division multiplexing (DWDM) backhaul links: Only passive components are needed in the field, no additional trunk fibers are required, and transceivers can flexibly be used for different services [2-4].

Coherent transmission has advantages over Intensity Modulation-Direct Detect (IM-DD) technology, which has been used up to the recent ITU-T 50G-PON [5], that make it particularly beneficial for the proposed application: A larger link budget results from the increase in the receiver sensitivity by coherent detection and a lower fiber attenuation in the C-/L-band compared to the O-band [6]. Digital signal processing compensates signal distortions caused by chromatic dispersion. The higher and scalable data rates of coherent PON (CPON) systems will provide new opportunities for optical network convergence (e.g. PON access and mobile x-hauling) and for addressing premium business applications.

Network operators today maintain separate backhaul and access networks, the former commonly based on multiplexers/demultiplexers (MUX/DEMUXs) with 100 GHz grid optical channels. To ensure coexistence over brownfield networks it is crucial to investigate the system architectures and transceiver specifications that allow the operation of 100 Gbit/s CPON.

PON transceivers have much stricter cost constraints than their DWDM transport counterparts in which a full-band tuneable laser with accurate wavelength control contributes to a large part of the transceiver cost [7]. A relaxed laser frequency accuracy [8] may be a major market enabler for CPON products. However, this may worsen the transmission performance when propagating the signals over a fixed DWDM filter grid.

In this contribution, we numerically evaluate the performance of different architectures for CPON signal transmission over brownfield DWDM links. Using measured profiles of field deployed DWDM multiplexers and demultiplexers (MUX/DEMUX), we identify the best architecture allowing a relaxation of the laser frequency accuracy.

**System architecture**
The CPON systems could coexist with and operate over a brownfield DWDM line system using passive DWDM devices, such as MUX/DEMUXs. Two potential architectures that enable coexistence are shown in Fig. 1. Both architectures comprise a conventional splitter-based optical distribution network (ODN) which follows a DWDM trunk link employing MUX/DEMUX filters.
The architecture shown in Fig. 1a) relies on having separate DWDM channels for the upstream (US) and downstream (DS) signals. If the channels have sufficient frequency separation, inexpensive diplexers (DXs) can be used to separate US and DS directions. This option will be referred to as *dual-channel architecture*, whereas the architecture shown in Fig. 1b) contains both the US and the DS signals in a single DWDM channel and relies on optical circulators to physically separate them at the OLT/ONU. This option will be referred to as the *single-channel architecture*. The two options offer different advantages and disadvantages: most notably the single-channel architecture doubles the spectral efficiency by halving the number of DWDM channels required per bidirectional service, however this halves the available bandwidth for each transmission direction over a fixed optical filter, limiting future data

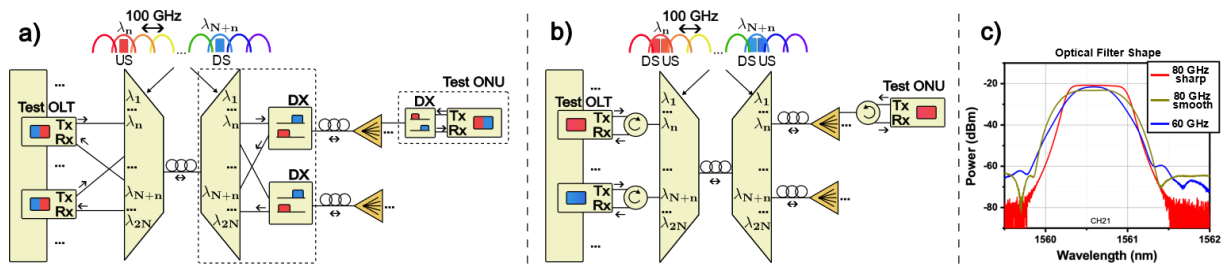

**Fig. 1:** Considered system architectures for PON and WDM coexistence. a) Dual-channel architecture with diplexers to isolate US and DS signals. b) Single-channel architecture with circulators to separate US and DS signals. c) Measured optical filters profiles of WDM MUX/DEMUXs

rate scalability. Finally, DXs can be easily integrated as part of the transceivers or MUX/DEMUXs while optical circulators are fiber-based components whose integration might prove challenging, thus increasing the overall transceiver cost which is critical for competitive ONUs.

In the case of the single-channel architecture, two different transceiver configurations are considered depending on the number of lasers inside the module: 1) two-laser configuration, where one is for US and the other is for DS; 2) single-laser configuration, where modulation around a digital carrier in the lower or the upper sideband (LSB or USB) around the laser frequency is used for DS and US, respectively. The two configurations have both merits and drawbacks: a single laser may deliver lower optical power than two separate lasers and modulation of the signal in a single sideband requires twice the bandwidth of electronic and optoelectronic components. Two separate lasers may be challenging to integrate and may drift relative to each other increasing the required frequency guard-band between US and DS signals.

**Simulation Model**
To evaluate the performance of the considered system scenarios, we build a simulation setup in VPIphotonics Design Suite 11.4 replicating the architectures shown in Fig. 1.

As depicted in Fig. 1 (c), we consider three different measured MUX/DEMUX filter profiles operating on a 100 GHz grid, specifically a Gaussian shape with ≈60 GHz -3dB bandwidth and two flat top shapes trading-off in-band flatness and filter steepness with ≈80 GHz -3dB bandwidth. For all components we use the same insertion loss of 3.5 dB to provide equal performance and focus the analysis on the filtering effect. However, in practical implementations wideband flat top designs may incur additional losses. The DWDM section of the system comprises a 30 km-long bidirectional standard single-mode fiber (SSMF) with an attenuation coefficient of 0.2 dB/km, dispersion coefficient of 16.3 ps/nm/km at 1550 nm, and a dispersion slope of 0.056 ps/nm2/km. Over the DWDM link, three signals with equal spectral characteristics propagate in adjacent 100 GHz slots and we consider the middle channel at $\lambda_n$ = 1560.06 nm for the performance estimation. The conventional PON link is modelled as a 20 km long linear lossless SSMF and a variable optical attenuator accounting for the available PON power budget and allocating for the loss of the fiber and of the splitter.

Both DXs and circulators have 1 dB of insertion loss. We neglect additional filtering effects over the channel of interest for the DX and we assume -30 dB isolation between circulator ports. We take this parameter into account to include in the simulator the impact of reflections between the Tx and Rx ports of the OLT/ONU. Rayleigh backscattering in the fiber is not simulated as the impact of the circulator is expected to be the main contribution to the overall optical return loss of the link, which usually has tighter requirements than -30 dB [5].

The transceivers operate with 30 Gbaud 100 Gbit/s quadrature phase shift keying (QPSK) signals and achieve an overall link budget of 38 dB with a launch power of 0 dBm at the Tx side and a receiver sensitivity of -38 dBm at 2% bit-to-error ratio (BER) before forward error correction (FEC). This operating point is based on previous experiments, where -40 dBm sensitivity was demonstrated using current components [9], and on efficient FEC implementations used in coherent pluggable modules following the OpenZR+ multi-source agreement [10]. Two different transceiver configurations are considered depending on one or two laser diodes assumed inside the module: 1) Two-laser configurations provide 16 dBm power per laser with a linewidth of 1 MHz, modulate a baseband signal with a 10% roll-off factor, and assume -3 dB analog bandwidth of 21 and 22.5 GHz for Tx and Rx, respectively. 2) The one-laser configuration assumes the same laser parameters except 19 dBm power equally split between Tx and Rx. USB or LSB modulation with a digital carrier at 16.5 GHz spacing from the laser frequency is set to avoid interference due to reflections between US and DS signals. The analog bandwidth of the devices for this configuration is doubled to provide the same baseline performance. Realistic modulation losses are set, resulting in output power between -8 and -9 dBm depending on the modulation scheme (baseband vs modulation over a digital carrier). A booster amplifier is used at the Tx to enforce 0 dBm launch power in all cases. The device is characterized by a noise figure of 7 dB.

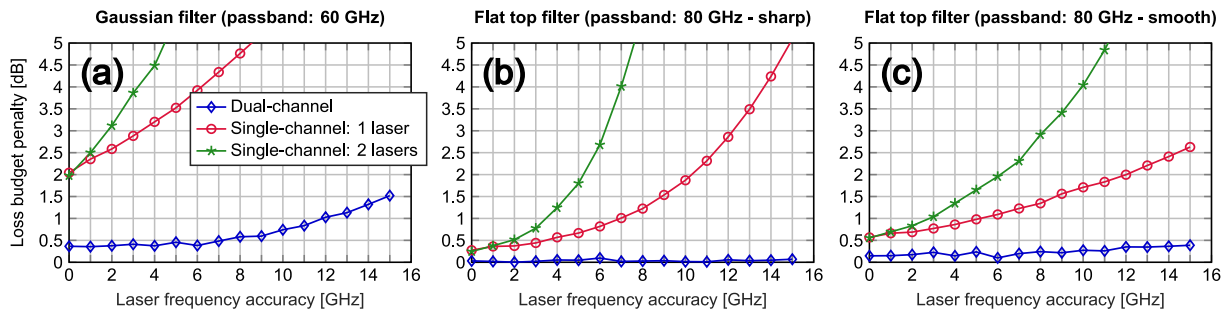

**Figure 2:** Numerical simulation results for the three considered optical filter shapes.

Finally, the receiver digitizes the received signals with 6 bits resolution at 2 samples/symbol and the digital signal processing chain includes signal orthonormalization, chromatic dispersion compensation, static matched filtering (including intermediate frequency demodulation for the single-laser case), 2x2 multi-input-multi-output equalization, downsampling, and carrier phase recovery (CPR). The CPR exploits the knowledge of pilot symbols every 32 payload symbols to realistically mitigate cycle slips [10]. Pilot data are discarded before error counting over ≈ 500k bits.

**Results**

For the performance estimation, downstream transmission is considered. The central frequencies of the OLT and ONU lasers are set equal, assuming accurate locking capability of the ONU to the OLT [11]. However, they are jointly swept to study the filtering penalty versus the laser frequency accuracy. The nominal frequency coincides with the center of the MUX/DEMUX optical passband for the dual-channel architecture and the single-channel architecture using a single laser. For the single-channel, two-lasers scenario, the lasers are shifted to ±16.5 GHz plus the assumed laser frequency accuracy ($f_{acc}$) around the center of the filter passband. This ensures US and DS signals do not spectrally overlap, avoiding significant reflection penalties.

The performance is measured in terms of the observed reduction in the PON loss budget for which the 2% BER threshold is reached compared to the case in which no filtering and reflection penalties are present. The results are shown in Fig. 2 for the three optical filter shapes. We observe that for the worst-case Gaussian filter (Fig. 2 (a)) only the dual-channel architecture is able to preserve acceptable performance, with less than 1 dB loss budget penalty even assuming $f_{acc} = 12$ GHz. Due to the narrow passband, both scenarios for the single-channel architecture exhibit instead a penalty of almost 2 dB even considering perfect laser locking. The results are improved for the flat top filters as can be seen from Figs. 2 (b) and (c). For the "80 GHz - sharp" filter, no visible penalty is observed for the dual-channel architecture for $f_{acc}$ < 15 GHz. On the counter side, this filter is characterized by a sharp transition to the stop-band. As a result, for the single-channel architectures large penalties > 1 dB are observed for $f_{acc}$ > 3.5 GHz and > 7 GHz in the single-laser and two-lasers scenarios, respectively. The "80 GHz - smooth" filter exhibits slightly worse in-band flatness, leading to < 0.5 dB link budget penalty for the dual-channel architecture across the studied range. However, the smoother transition band relaxes the requirements in the single-channel scenarios. In particular, < 1 dB penalty is obtained up to $f_{acc} = 6$ GHz.

Finally, a small penalty of ≈ 0.3 dB is observed with ideal laser accuracy and almost perfect filter flatness between the dual-channel and single-channel architectures. This is due to the presence of the reflection at the circulator, which contributes to the saturation of the receiver.

**Conclusions**

We numerically evaluate the performance of different architectures for 100 Gbit/s CPON coexistence over brownfield DWDM links accounting for filtering limitations of measured deployed DWDM MUX/DEMUXs. Arranging US and DS signals over two DWDM channels with sufficient spectral separation promises to lead to a cost-effective system architecture. It causes < 1 dB penalty over a 38 dB link budget even considering worst-case filtering with 60 GHz -3 dB Gaussian optical passband and a relaxed laser accuracy of up to 12 GHz. The architecture enables the use of simpler laser wavelength control compared to conventional transceivers for DWDM transport.

**Acknowledgements**

We would like to thank Steve Jia at CableLabs for providing the characteristics of field-deployed filters (Fig. 1c) and fruitful discussion.

This work has been partially funded by the German Federal Ministry of Economics and Climate Action in the project Helios-KT (16IPCEI201).

**References**

[1] ITU-T Standardization Q2/15 – Optical system for fibre access networks, 2022-2024, https://www.itu.int/net4/ITU-T/lists/q-text.aspx?Group=15&Period=17&QNo=2&Lang=en

[2] P. Ossieur, et al. "Demonstration of a 32× 512 split, 100 km reach, 2× 32× 10 Gb/s hybrid DWDM-TDMA PON using tunable external cavity lasers in the ONUs." Journal of Lightwave Technology, 2011.

[3] CableLabs, "Coherent Passive Optical Networks 100 Gbps Single-Wavelength PON," CableLabs Specifications, May 3rd 2023, https://www.cablelabs.com/specifications/CPON-SP-ARCH accessed on 24 April 2024.

[4] Z. Jia, et.al., "Revolutionizing Access Networks: How TFDM Coherent PON Combines the Best of TDM and WDM PONs", NCTA technical papers, 2023.

[5] "50-Gigabit-capable passive optical networks (50G-PON): Physical media dependent (PMD) layer specification," ITU-T Recommendation G.9804.3, 2021, https://www.itu.int/rec/T-REC-G.9804.3/en

[6] "Optical interfaces for coarse wavelength division multiplexing applications," ITU Recommendation ITU-T G.695, 2018, https://www.itu.int/rec/T-REC-G.695/en

[7] J. Müller, O. Jovanovic, T. Fehenberger, G. Di Rosa, J.P. Elbers, and C. Mas-Machuca, "Multi-wavelength transponders for high-capacity optical networks: a physical-layer-aware network planning study," Journal of Optical Communication and Networking 15, 2023.

[8] Y. Luo, H. Roberts, K. Grobe, M. Valvo, D. Nesset, K. Asaka, H. Rohde, J. Smith, J. Wey, and F. Effenberger, "Physical Layer Aspects of NG-PON2 Standards—Part 2: System Design and Technology Feasibility," Journal of Optical Communication and Networking, 2016.

[9] G. R. Martella, A. Nespola, S. Straullu, F. Forghieri, and R. Gaudino, "Scaling laws for unamplified coherent transmission in next-generation short-reach and access networks," Journal of Lightwave Technology 39, 5805–5814, 2021, DOI: 10.1109/JLT.2021.3092523.

[10] "OpenZR+ Multi-source agreement Technical Specification", [online] Available: http://openzrplus.org/documents/.

[11] S. Ranzini *et al.*, "Local and remote laser frequency control in point-to-multipoint networks using digital subcarriers," *49th European Conference on Optical Communications (ECOC 2023)*, Hybrid Conference, Glasgow, UK, 2023, pp. 807-810, doi: 10.1049/icp.2023.2344.